\documentclass[12pt]{article}

\DeclareTextFontCommand{\textwasy}{\wasyfamily}
\def \wasyfamily{\fontencoding{U}\fontfamily{wasy}\selectfont}

\DeclareTextCommand{\dh}{OT1}{{\wasyfamily\char107}}

\newcommand{\eth}{{\textrm{\dh}}}

 \def \bea { \begin{eqnarray}}
\def \eea {\end{eqnarray}}
\def \be {\begin{equation}}
\def \ee {\end{equation}}

\newcommand{\mb}[1]{{m_{(#1)}}}

\newcommand{\M}[1]{{\stackrel{#1}{M}}}  % In iopart class, this should read {{\overset{#1}{M}}}

\newcommand{\ba}{\bar{a}}

\title{On algebraically special vacuum spacetimes in five dimensions}
\author{Harvey S. Reall\footnote{hsr1000@cam.ac.uk}, Alexander A.H. Graham\footnote{aahg2@cam.ac.uk} and Carl P. Turner\footnote{cpt39@cam.ac.uk} \\{\small Department of Applied Mathematics and Theoretical Physics, University of Cambridge,} \\ {\small Centre for Mathematical Sciences, Wilberforce Road, Cambridge CB3 0WA, UK}}

\begin{document}

\maketitle

\begin{abstract}
Vacuum solutions admitting a hypersurface-orthogonal repeated principal null direction are an important class of 4d algebraically special spacetimes. We investigate the 5d analogues of such solutions: vacuum spacetimes admitting a hypersurface-orthogonal multiple Weyl aligned null direction (WAND). Such spacetimes fall into 4 families determined by the rank of the $3 \times 3$ matrix that defines the expansion and shear of the multiple WAND. The rank 3 and rank 0 cases have been studied previously. We investigate the 2 remaining families. We show how to define coordinates which lead to a considerable simplification of the Einstein equation with cosmological constant. The rank 2 case gives warped product and Kaluza-Klein versions of the 4d Robinson-Trautman solutions as well as some new solutions. The rank 1 case gives product, or analytically continued Schwarzschild, spacetimes.
\end{abstract}

\section{Introduction}

The technique for solving the 4d Einstein equation that has received most attention is searching for solutions with algebraically special Weyl tensor \cite{exact}. This is how the Kerr solution was discovered. It is natural to exploit the same technique to solve the Einstein equation for $d>4$ spacetime dimensions, especially as this is one of the few techniques which works with a cosmological constant. This paper will consider only Einstein spacetimes, i.e. solutions of the vacuum Einstein equation with a cosmological constant. We will look for solutions which are algebraically special in the classification of Coley, Milson, Pravda and Pravdova \cite{Coley:2004jv}. Such a solution admits a {\it multiple Weyl aligned null direction} (multiple WAND), the higher-dimensional generalization of a repeated principal null direction.

In 4d, the Einstein equation simplifies considerably when the repeated principal null direction is hypersurface-orthogonal. There are two families of such solutions: the Robinson-Trautman (RT) family and the Kundt family \cite{exact}. RT spacetimes are defined by the property of admitting a null geodesic congruence with vanishing shear and rotation, but non-vanishing expansion. This family includes the Schwarzschild solution, the C-metric, and time-dependent spacetimes which approach Schwarzschild asymptotically. Kundt spacetimes are defined by the property of admitting a null geodesic congruence with vanishing expansion, rotation and shear. This family includes pp-waves, and the near-horizon geometry of the extreme Kerr black hole.

Higher-dimensional RT spacetimes were investigated in Ref. \cite{Podolsky:2006du}. It was found that these are significantly less rich than in 4d, containing certain simple time-independent generalizations of the Schwarzschild solution but nothing that could be identified as a higher-dimensional analogue of the C-metric, and no time-dependent generalizations of Schwarzschild. Higher-dimensional Kundt spacetimes were discussed in Refs. \cite{Coley:2002ku,Podolsky:2008ec}. 

A possible reason why the higher-dimensional RT solutions are less interesting than in 4d is that a multiple WAND in $d>4$ dimensions need not be shear-free. Hence, unlike in 4d, the RT and Kundt families are not the most general solutions admitting a hypersurface orthogonal multiple WAND. For example, a Schwarzschild black string belongs to neither family. So it seems worthwhile investigating more general solutions admitting a hypersurface-orthogonal multiple WAND. Ref. \cite{Pravdova:2008gp} studied some general properties of such solutions. Coordinates were introduced analogous to those used for RT or Kundt solutions. The dependence of the metric on one of these coordinates was fully determined for the more special algebraic types.

Recently, Ref. \cite{Ortaggio:2012hc} investigated algebraically special solutions in 5d. In the case of a hypersurface-orthogonal multiple WAND, it was found that there exist two classes of spacetime that are distinct from the RT and Kundt classes. To describe these classes, we need to introduce some notation. Recall that the expansion, rotation and shear of a null vector field $\ell^a$ are defined as follows. Introduce a set of $d-2$ orthonormal spacelike vectors $m_i^a$, orthogonal to $\ell^a$. Define the $(d-2) \times (d-2)$ {\it optical matrix} of $\ell^a$
\be
\label{rhodef}
 \rho_{ij} = m_i^a m_j^b \nabla_b \ell_a
\ee
The expansion, rotation and shear are the trace, antisymmetric part, and traceless symmetric part of $\rho_{ij}$. 
We are interested in the case of a hypersurface orthogonal multiple WAND $\ell^a$, for which $\rho_{ij}$ is symmetric. In this case, in 5d, Ref. \cite{Ortaggio:2012hc} found that the eigenvalues of the $3 \times 3$ matrix $\rho_{ij}$ are either $\{a,a,a\}$, $\{a,a,0\}$, $\{a,0,0\}$ or $\{0,0,0\}$, where $a \ne 0$. The first of these is the RT family and the last is Kundt family. We will investigate the two other possibilities. Both classes are non-empty e.g. they contain the Schwarzschild black string solution, and the product $dS_3 \times S^2$ respectively.

We proceed as follows. The first step is to introduce coordinates adapted to the geometrical properties of these spacetimes. For RT or Kundt spacetimes, the procedure for doing this is well-known. However, in the cases of interest to us we have to work harder. In section \ref{sec:coords} we show that these spacetimes admit foliations by null submanifolds which enable one to define a natural set of coordinates adapted to the structure of $\rho_{ij}$. We then investigate the vacuum Einstein equation, allowing for a cosmological constant. 

In section \ref{sec:rank2} we consider the case for which $\rho_{ij} $ has rank 2. The general solution involves a function $m$ similar to the "mass" parameter of the 4d RT solutions. For $m \ne 0$ we find that the solution must belong to one of the following classes (i) for any cosmological constant $\Lambda$, a warped product involving a 4d vacuum RT solution (equation (\ref{RTwarp})); (ii) for $\Lambda=0$, the Kaluza-Klein oxidation of a 4d RT solution with a null Maxwell field (equation (\ref{KK})); (iii) for $\Lambda =0$, a new class with metric given in equation (\ref{newmetric1}), for which the Einstein equation reduces to a pair of elliptic PDEs in 2 dimensions; (iv) for $\Lambda<0$ a similar class, with metric given in (\ref{newmetric2}), involving the solutions of PDEs in 3 dimensions.  For $m=0$ the Einstein equation reduces to a pair of equations identical to those governing type III (or more special) RT solutions of 4d Einstein-Maxwell theory (section \ref{sec:RTzerom}). 

When $\rho_{ij}$ has rank 1, we determine all solutions explicitly in section \ref{sec:rank1}. They coincide with the solutions obtained in Ref. \cite{Durkee:2009nm}. They are either product spacetimes ($dS_3 \times S^2$ or $adS_3 \times H^2$) or analytically continued versions of a generalized Schwarzschild solution (equation (\ref{kkbubble})). Examples of the latter are the Kaluza-Klein bubble of Ref. \cite{Witten:1981gj} and the anti-de Sitter soliton of Ref. \cite{Horowitz:1998ha}.

Finally, we discuss a 6d case. Ref. \cite{Ortaggio:2012cp} obtained necessary conditions on the eigenvalues of the optical matrix of a hypersurface-orthogonal multiple WAND in any number of dimensions. In 6d, there is the interesting possibility of eigenvalues $\{a,a,b,b\}$ with $a,b \ne 0$, $a \ne b$. The coordinates of section \ref{sec:coords} can be introduced for this case. However, the only solutions of the Einstein equation (with cosmological constant) in this class are conformally flat.

\section{Canonical form for metric}

\label{sec:coords}

\subsection{Introduction of coordinates}

We will make use of the higher-dimensional generalization of the Geroch-Held-Penrose (GHP) formalism introduced in Ref. \cite{Durkee:2010xq}. Our notation will follow the notation of that reference. Introduce a null basis $\{ e_0 \equiv  \ell,e_1 \equiv n, e_i \equiv m_i \}$, $i = 2, \ldots ,d-1$ satisfying
\be
 \ell^2 = n^2 = \ell \cdot m_i = n \cdot m_i = 0 \qquad \ell \cdot n = 1 \qquad m_i \cdot m_j = \delta_{ij}
\ee
Greek indices $\mu,\nu, \ldots$ refer to this basis. Latin indices $a,b,c\ldots$ are abstract indices. Introduce the following notation for the connection components \cite{Pravdaetal04}
\be
 L_{\mu\nu} = \nabla_\nu \ell_\mu, \qquad N_{\mu\nu} = \nabla_\nu n_\mu, \qquad \stackrel{i}{M}_{\mu\nu} = \nabla_\nu  (m_i)_\mu
\ee
Consider an Einstein spacetime, in any number of dimensions, admitting a multiple WAND. Ref. \cite{Durkee:2009nm} proved that such a spacetime must admit a {\it geodesic} multiple WAND so there is no loss of generality in restricting attention to geodesic multiple WANDs. We start with a couple of useful Lemmas. Note that we do not yet assume our multiple WAND to be hypersurface-orthogonal. 

\medskip

\noindent {\bf Lemma 1}. Let $\ell^a$ be a geodesic multiple WAND (not necessarily affinely parameterized) such that (i) $\rho_{ij}$ (defined in (\ref{rhodef})) has a block diagonal structure with blocks $\rho_{IJ}$ and $\rho_{\alpha \beta}$, i.e., $\rho_{I \alpha} = \rho_{\alpha I} = 0$; (ii) $\rho_{IJ} = \lambda \delta_{IJ}$ for some function $\lambda$; (iii) $\lambda$ is not an eigenvalue of $\rho_{\alpha \beta}$. Then the distribution ${\cal D} \equiv {\rm span}\{ \ell,m_I \; (\forall I) \}$ is integrable.
 
 \medskip 
 
\noindent {\it Proof.} Using (i), equation NP1 of Ref. \cite{Durkee:2010xq} with $i=\alpha$, $j=I$ gives
\be
 \rho_{\alpha \beta} \stackrel{\beta}{M}_{I0} =  \rho_{JI} \stackrel{\alpha}{M}_{J0} = \lambda \stackrel{\alpha}{M}_{I0} 
\ee
where we used (ii) in the final equality. Now (iii) implies that $\stackrel{\alpha}{M}_{J0} = 0$, i.e.,
\be
\label{partrans}
 m_J \cdot \left( \ell \cdot \nabla m_\alpha \right) = 0
\ee  
Consider $[\ell, m_i]$. First we have $\ell \cdot [\ell, m_i] = \ell^a \ell^b \nabla_b m_{ia} - \ell^a m_i^b \nabla_b \ell_a = - m_{ia} \ell^b \nabla_b \ell^a - (1/2) m_i^b \nabla_b \ell^2 = 0$ using the geodesic equation for $\ell^a$. We also have $m_{\alpha} \cdot [\ell, m_I] =m_\alpha \cdot ( \ell \cdot \nabla m_I) - m_\alpha \cdot ( m_I \cdot \nabla \ell) = -m_I  \cdot (\ell \cdot \nabla m_\alpha) - \rho_{\alpha I} =  0$. These results imply $[\ell, m_I] \in {\cal D}$.  

Using (i) and (ii), equation NP3 of Ref. \cite{Durkee:2010xq} with $i=\alpha$, $j=J$, $k=K$ reduces to
\be
 \rho_{\alpha \beta}  \stackrel{\beta}{M}_{[JK]} = \lambda \stackrel{\alpha}{M}_{[JK]} 
\ee
and hence (iii) implies $\stackrel{\alpha}{M}_{[JK]} =0$, i.e., 
\be
  m_{[J} \cdot \left( m_{K]} \cdot \nabla m_\alpha \right) = 0
\ee
Finally, consider $[m_J,m_K]$. We have $\ell \cdot [m_J, m_K] = - 2m_{[K} \cdot ( m_{J]} \cdot \nabla \ell) = -2 \rho_{KJ} = 0$ using (ii). We also have $m^\alpha \cdot [m_J,m_K] = - 2m_{[K} \cdot ( m_{J]} \cdot \nabla m_\alpha) = 0$ using the result just obtained. These results imply $[m_J,m_K] \in {\cal D}$. This concludes the proof. 

\medskip

\noindent {\bf Lemma 2}. The basis vectors $m_i$ can be chosen to be parallelly transported along geodesics with tangent $\ell^a$ whilst preserving the properties (i), (ii), (iii) of Lemma 1. 
\medskip

\noindent {\it Proof}.  Let $V^a$ be parallelly transported $\ell^a \nabla_a V^b =0 $ with $V^1 = V \cdot \ell = 0$ (a condition preserved by parallel transport). Then
\be
 \ell^a \partial_a V_i = \ell^a \nabla_a \left( V \cdot m_i \right) = V^\mu   \stackrel{i}{M}_{\mu 0} = V^0  \stackrel{i}{M}_{0 0} + V_j  \stackrel{i}{M}_{j 0} = V_{j}  \stackrel{i}{M}_{j 0} 
\ee 
 where, in the final equality, we used  $\stackrel{i}{M}_{0 0} = \ell_a \ell^b \nabla_b m_i^a = - m_i^a\ell^b \nabla_b \ell_a = 0$ (from the geodesic equation). Using (\ref{partrans}) (i.e. $\stackrel{\alpha}{M}_{J0} = 0$) we have
 \be
  \ell \cdot \partial V_I = V_J \stackrel{I}{M}_{J0}, \qquad \ell \cdot \partial V_\alpha = V_\beta \stackrel{\alpha}{M}_{\beta 0} 
 \ee
Hence if all $V_I$ (all $V_\alpha$) initially are zero then all $V_I$ (all $V_\alpha$) remain zero under parallel transport. Now define $m_i'$ to be a set of basis vectors which coincide initially with $m_i$ and are parallelly transported along a geodesic with tangent $\ell$. Using the result just derived, $m_I'$ remains orthogonal to $m_\alpha$ and $m_\alpha'$ remains orthogonal to $m_I$ (and both remain orthogonal to $\ell$). It follows that $m_I'$ is a linear combination of $\ell$ and the $m_J$, and $m_\alpha'$ is a linear combination of $\ell$ and the $m_\beta$. Hence the new basis is related to the old one by a null rotation \cite{Durkee:2010xq} about $\ell$ (which does not change $\rho_{ij}$) and a spin \cite{Durkee:2010xq} which does not mix $m_I$ with $m_\alpha$ and hence preserves the structure of $\rho_{ij}$. 

\medskip

In the rest of this paper, we will impose the extra condition that the geodesic multiple WAND $\ell^a$ is hypersurface-orthogonal, which is equivalent to vanishing rotation:
\be
 \rho_{[ij]} = 0.
\ee
Let $\rho_\iota$, $\iota=1, \ldots, N$ be the distinct eigenvalues of $\rho_{ij} = \rho_{(ij)}$. Choose the spatial basis vectors $\{ m_i \}$ = $\{ \mb{I_1}, \ldots ,\mb{I_N} \}$ so that $\{ \mb{I_\iota} \}$ spans the $\iota$th eigenspace of $\rho_{ij}$. Then $\rho_{(ij)}$ is diagonal. Lemma 1 implies that the distribution ${\cal D}_\iota \equiv \{ \ell, m_{I_\iota} \}$ is integrable for all $\iota$. Therefore each eigenspace of $\rho_{ij}$ determines a foliation by null submanifolds ${\cal N}_\iota$ of spacetime.

 Let us focus on the case in which $N=2$, i.e., $\rho_{ij}$ has two distinct eigenvalues. For $d=5$ this is the only possibility other than Robinson-Trautman or Kundt  \cite{Ortaggio:2012hc} (we also know that $\rho_2=0$ in this case although we shall not assume this yet). We can introduce coordinates adapted to these null foliations as follows.

$\ell$ is hypersurface orthogonal. Introduce a function $u$ with $du \ne 0$ so that $\ell$ is orthogonal to surfaces of constant $u$. Using a boost \cite{Durkee:2010xq} we can arrange that $\ell_a = -(du)_a$, which implies that $\ell$ is tangent to the affinely parameterized null geodesic generators of the surfaces of constant $u$. 

Let the surfaces ${\cal N}_1$ have dimension $n+1$ and introduce a chart 
\be (\tilde{X}^0,\ldots, \tilde{X}^{n},\tilde{Y}^{n+1}, \ldots, \tilde{Y}^{d-1})
\ee
 so that the surfaces ${\cal N}_1$ are surfaces of constant $\tilde{Y}$. These lie within the surfaces of constant $u$ therefore $u$ is constant over each surface ${\cal N}_1$ and hence cannot depend on $\tilde{X}$. Therefore $u=u(\tilde{Y})$. Hence we can define new coordinates $(\tilde{X},y^{n+1},\ldots,y^{d-2},u)$ where each $y$ is a function of the $\tilde{Y}$s and the surfaces ${\cal N}_1$ are surfaces of constant $y^{n+1},\ldots, y^{d-2},u$. Next pick some spacelike hypersurface $\Sigma$ transverse to $\ell$ and complete $(y^{n+1}, \ldots, y^{d-2} ,u)$ to a chart $(y^1, \ldots y^{d-2},u)$ on this surface. Finally, assign coordinates $(u,r,y^1, \ldots, y^{d-2})$ to the point affine parameter distance $r$ from this surface along the integral curve of $\ell$ (a geodesic) starting from $(y^1, \ldots y^{d-2},u)$. Since we already know that $u$ and $y^{n+1}, \ldots y^{d-2}$ are constant along such geodesics, this agrees with our previous definitions of these quantities. We now have a chart $(u,r,y^1, \ldots y^{d-2})$ for which $\ell^a = \partial/\partial r$, $\ell_a = -(du)_a$, and the surfaces ${\cal N}_1$ are surfaces of constant $u,y^{n+1},\ldots, y^{d-2}$. On any such surface, we have a coordinate chart defined by $(r,y^1,\ldots ,y^n)$.
 
Repeat this procedure for the foliation defined by ${\cal N}_2$, using the same surface $\Sigma$. This gives a coordinate chart $(u,r,x^1, \ldots, x^{d-2})$  with the property that the surfaces ${\cal N}_2$ are surfaces of constant $u,x^1, \ldots, x^n$ (after reordering the $x$-coordinates). The coordinates $u,r$ are common to both charts.
  
Consider $\eta = dx^1 \wedge \ldots \wedge dx^n \wedge dy^{n+1} \wedge \ldots \wedge dy^{d-2}$. If $\eta=0$ then some linear combination of $dx^1, \ldots , dx^n$ equals a linear combination of $dy^{n+1}, \ldots , dy^{d-2}$. But then there is some 1-form other than $(du)_a = -\ell_a$ that is normal to both ${\cal N}_1$ and ${\cal N}_2$, which is a contradiction. Hence $\eta \ne 0$. 

Now $\ell = \partial/\partial r$ (in either chart) so $\ell \cdot dy^\mu = \ell \cdot dx^\mu = 0$ hence $\ell \cdot \eta=0$. Therefore $\ell \cdot (dr \wedge \eta) = \eta \ne 0$ and hence $dr \wedge \eta \ne 0$. 
It follows that $(r,x^1, \ldots, x^n, y^{n+1}, \ldots, y^{d-2})$ is a good coordinate chart on surfaces of constant $u$ and then $(u,r,x^1, \ldots, x^n, y^{n+1}, \ldots, y^{d-2})$ is a good coordinate chart on spacetime with the property that surfaces ${\cal N}_1$ are surfaces of constant $u,y^{n+1}, \ldots, y^{d-2}$ and ${\cal N}_2$ are surfaces of constant $u,x^1, \ldots, x^n$. The distribution spanned by $\{\partial/\partial r, \partial/\partial x^1, \ldots ,\partial/\partial x^n\}$ is tangent to ${\cal N}_1$ and the distribution spanned by $\{\partial/\partial r, \partial/\partial y^{n+1}, \ldots ,\partial/\partial y^{d-2} \}$ is tangent to ${\cal N}_2$. Hence these distributions are orthogonal. 

The metric now must take the form
\bea
\label{metric}
 ds^2&=& -g^{rr} du^2 - 2 du dr + g_{MN} \left( dx^M + g^{rM} du \right) \left(  dx^N + g^{rN} du \right) \nonumber \\ &+&  g_{AB} \left( dy^A + g^{rA} du \right)   \left( dy^B + g^{rB} du \right)  
\eea
where indices $M,N$ run from $1$ to $n$ and $A,B$ run from $n+1$ to $d-2$. This is a special case of the metric discussed in Ref. \cite{Pravdova:2008gp}. Henceforth we will use (as in the above Lemmas) indices $I,J,\ldots$ instead of $I_1, J_1, \ldots$ and $\alpha,\beta, \ldots$ instead of $I_2, J_2, \ldots$. We will use $m,n,\ldots$ to denote either $M$ or $A$. Note that the basis vectors $m_i$ obey $m_I^u=m_\alpha^u=m_I^A=m_\alpha^M=0$. 

\subsection{Further simplifications}

We will show how the above metric can be simplified further by imposing the multiple WAND condition and some components of the Einstein equation. We will do this by solving some of the "Newman-Penrose" (NP) equations of Ref. \cite{Durkee:2010xq}.

We have chosen $\ell^a$ to be tangent to affinely parameterized geodesics, and will take $m_i^a$ to be parallely transported along these geodesics. It follows that $n^a$ also must be parallely transported, i.e., all basis vectors are parallely transported. 

In our parallely transported basis, equation NP1 of Ref.  \cite{Durkee:2010xq} (with the WAND condition) after shifting $r$ appropriately reduces to 
\be
 \rho_1 = \frac{1}{r}, \qquad \rho_2 = \frac{1}{r+\bar{a}(u,x,y)}
\ee
where $\bar{a}(u,x,y) \ne 0$ and we will allow for the possibility $\bar{a}=\infty$ corresponding to $\rho_2 = 0$ (we know $\bar{a}=\infty$ when $d=5$). All equations below are valid in the limit $\bar{a} \rightarrow \infty$ unless otherwise noted. Here, and henceforth, an overbar denotes a quantity independent of $r$. 

Calculating $\rho_{ij}$ gives 
\be
 \rho_{ij} = \frac{1}{2} m_i^m m_j^n \partial_r g_{mn}
\ee
and hence
\be
\label{drg}
 \frac{1}{2} \partial_r g_{mn} = \rho_{ij} m_{im} m_{jn}
\ee
Taking  $\mu=m$ in $\ell \cdot \nabla  m_i^\mu=0$ and using (\ref{drg}) gives an equation which can be solved to give
\be
 m_I^m(u,r,x,y) = \frac{1}{r} \bar{m}_I^m(u,x,y), \qquad m_\alpha^m (u,r,x,y) = \frac{1}{1+r/\bar{a}} \bar{m}_\alpha^m(u,x,y) 
\ee
Taking $\mu=r$ gives
\be
\label{ppcond}
 \partial_r g^{rm} = 2 (\partial_r m_i^r) m_i^m
\ee
We now have
\be
\label{inversemetric}
 g^{MN} = \frac{1}{r^2} \bar{m}_I^M \bar{m}_I^N, \qquad g^{AB} = \frac{1}{(1+r/\bar{a})^2} \bar{m}_\alpha^A \bar{m}_\alpha^B
\ee
and hence 
\be
 g_{MN} = r^2 \bar{g}_{MN}(u,x,y), \qquad g_{AB} = (1+r/\bar{a})^2 \bar{g}_{AB}(u,x,y)
\ee
Since $d\ell=0$ we have
\be
 L_{1j} = L_{j1} \equiv \tau_j
\ee
Equation NP3 of Ref. \cite{Durkee:2010xq} is
\be
\label{NP3}
 \eth_{[j} \rho_{|i|k]} = 0.
\ee
Setting $j=J,i=I,k=K$ gives
\be
\label{NP3a}
 \frac{1}{r} m_J^r + \tau_J = 0 \qquad {\rm if} \; n>1
\ee
(recall that $n$ is the number of $x$-coordinates, the restriction $n>1$ arises because (\ref{NP3}) is antisymmetric on $jk$). Setting $j=J,i=I,k=\gamma$ gives
\be
\label{MgammaIJ}
 \M{\gamma}_{IJ} = (1+r/\bar{a}) \left( \frac{1}{r} m_\gamma^r + \tau_\gamma \right) \delta_{IJ}
\ee
Setting $j=I,i=\alpha,k=\beta$ gives
\be
\label{MIalphabeta}
 \M{I}_{\alpha\beta} = - \frac{r}{\ba} \left[ \frac{1}{r+\ba} \left( m_I^r + m_I^M \partial_M \ba \right) + \tau_I \right] \delta_{\alpha \beta}
\ee
$j=\beta, I= \alpha, K=\gamma$ gives
\be
\label{NP3b}
 \frac{1}{(r+\ba)} \left( m_\alpha^r + m_\alpha^A \partial_A \ba \right) + \tau_\alpha = 0 \qquad {\rm if} \; d-2-n>1, \; \ba \ne \infty
\ee
The remaining components of (\ref{NP3}) are trivial. 

Since the basis is parallelly propagated, $\tau'_i \equiv N_{i0}=0$. Equation NP2 of Ref. \cite{Durkee:2010xq} (with vanishing curvature terms) reduces to
\be
 \tau_I = \frac{1}{r} \bar{\tau}_I(u,x,y), \qquad \tau_\alpha = \frac{1}{1+r/\ba}  \bar{\tau}_\alpha(u,x,y)
\ee
The orthogonality relations of the basis vectors give
\be
 n_r = 1, \qquad n_m = -m_i^r m_{im}
\ee
Using this we can calulate $\tau_i = L_{i1}$ directly to obtain
\be
 \tau_i = - \partial_r m_i^r + \frac{1}{2} m_{im} m_j^r m_{jn} \partial_r g^{mn}
\ee
Since we know the $r$-dependence of the LHS we now can obtain (using (\ref{inversemetric}))
\be
 m_I^r = - \bar{\tau}_I + \frac{1}{r} \bar{m}_I^r
\ee
\be
 m_\alpha^r = - \frac{r}{1+r/\ba} \bar{\tau}_\alpha + \frac{1}{1+r/\ba} \bar{m}_\alpha^r
\ee
Now using (\ref{NP3a}) gives
\be
\label{NP3a2}
 \bar{m}_I^r = 0  \qquad {\rm if} \; n>1
\ee
and (\ref{NP3b}) gives
\be
\label{NP3b2}
 \bar{m}_\alpha^r = - \ba \bar{\tau}_\alpha - \bar{m}_\alpha^A \partial_A \ba  \qquad {\rm if} \; d-2-n>1, \; \ba \ne \infty
\ee
A calculation from the definition of the LHS gives
\be
 \M{\alpha}_{IJ} = \frac{1}{1+r/\ba} \left( -\bar{\tau}_\alpha + \frac{1}{r} \bar{m}_{\alpha}^r \right) \delta_{IJ} - \frac{1}{2(1+r/\ba)} m_{IM} m_{JN} m_\alpha^A \partial_A g^{MN}
\ee
Comparing with (\ref{MgammaIJ}) gives
\be
 \partial_A \bar{g}_{MN} = 2 \bar{g}_{AB} \bar{m}_\alpha^B \left( \bar{\tau}_\alpha + \frac{1}{\ba} \bar{m}_\alpha^r \right) \bar{g}_{MN}
\ee
and hence
\be
 \bar{g}_{MN}(u,x,y) = \bar{P}(u,x,y)^{-2} h_{MN}(u,x)
\ee
where
\be
\label{Peq1}
 \frac{\partial_A \bar{P}}{\bar{P}} = -\bar{g}_{AB} \bar{m}_\alpha^B \left( \bar{\tau}_\alpha + \frac{1}{\ba} \bar{m}_\alpha^r \right)
\ee
Combining with (\ref{NP3b2}) gives (absorbing a function of $u,x$ into $h_{MN}$) 
\be
\label{Peq2}
 \bar{P}=\ba \qquad {\rm if} \; d-2-n>1, \; \ba \ne \infty
\ee
Now we calculate from the definition of the LHS
\be
 \M{I}_{\alpha\beta} = \frac{1}{r+\ba} \left( - \bar{\tau}_I + \frac{1}{r} \bar{m}_I^r \right) \delta_{\alpha\beta} - \frac{1}{2r} m_{\alpha A} m_{\beta B} \bar{m}_I^M \partial_M g^{AB}
\ee
Comparing with (\ref{MIalphabeta}) gives
\be
 \partial_M \bar{g}_{AB} = -\frac{2}{\ba} \bar{g}_{AB}\left(  \bar{g}_{MN} \bar{m}_I^N \bar{m}_I^r \right)
\ee
and hence
\be
 \bar{g}_{AB}(u,x,y) = \bar{Q}(u,x,y)^{-2} h_{AB}(u,y) 
\ee
where
\be
\label{Qeq1}
 \frac{\partial_M \bar{Q}}{\bar{Q}} =  \frac{1}{\ba} \bar{g}_{MN} \bar{m}_I^N \bar{m}_I^r
\ee
Combining with (\ref{NP3a2}) gives (absorbing a function of $u,y$ into $h_{AB}$)
\be
\label{Qeq2}
 \bar{Q}(u,x,y) = 1 \qquad {\rm if} \; \ba=\infty \; {\rm or} \; n>1
\ee
Now we return to (\ref{ppcond}). This gives
\be
 g^{rM} = \frac{1}{r^2} \bar{m}_I^r \bar{m}_I^M + \bar{g}^{rM} 
 \ee
so, from (\ref{NP3a2}),
\be
 g^{rM} = \bar{g}^{rM} \qquad {\rm if} \; n>1
\ee
and, from (\ref{Qeq1}),
\be
g^{rM} = \frac{\ba}{r^2} \bar{g}^{MN} \frac{\partial_N \bar{Q}}{\bar{Q}} + \bar{g}^{rM} \qquad {\rm if} \; \ba \ne \infty
\ee
From (\ref{ppcond}) we also obtain
\be
 g^{rA} = \frac{\ba}{(1+r/\ba)^2} \left( \bar{\tau}_\alpha + \frac{1}{\ba} \bar{m}_\alpha^r \right) \bar{m}_\alpha^A + \bar{g}^{rA} \qquad {\rm if} \; \ba \ne \infty
\ee
\be
 g^{rA} = -2r \bar{\tau}_\alpha \bar{m}_\alpha^A + \bar{g}^{rA} \qquad {\rm if} \; \ba = \infty
\ee
Note that the coefficient in the $r$-dependent piece involves (in both cases)
\be
 \left( \bar{\tau}_\alpha + \frac{1}{\ba} \bar{m}_\alpha^r \right) \bar{m}_\alpha^A = -\bar{g}^{AB} \partial_B \bar{P}/\bar{P}
\ee
In summary, we have shown that the metric components in (\ref{metric}) must take the following form:
\be
 g_{MN}(u,r,x,y) = r^2 \bar{P}(u,x,y)^{-2} h_{MN}(u,x)
\ee
where $\bar{P}=\bar{a}$ if $d-2-n>1$ and $\ba \ne \infty$
\be
 g_{AB}(u,r,x,y) =  (1+r/\bar{a}(u,x,y))^2\bar{Q}(u,x,y)^{-2} h_{AB}(u,y) 
\ee
where $\bar{Q}=1$ if $ \ba=\infty$ or $n>1$,
\be
 g^{rM}(u,r,x,y) = \frac{1}{r^2} \bar{f}^{rM}(u,x,y) + \bar{g}^{rM}(u,x,y), 
\ee
where $\bar{f}^{rM}=0$ if $n>1$ and
\be
 \bar{f}^{rM} = \frac{\ba}{r^2} \bar{P}^2 h^{MN} \frac{\partial_N \bar{Q}}{ \bar{Q}} \qquad {\rm if} \; \ba \ne \infty
\ee
($\bar{f}^{rM}$ is unconstrained if $n=1$, $\ba=\infty$.)
\be
 g^{rA} = -  \frac{\ba}{(1+r/\ba)^2} \bar{Q}^2h^{AB} \frac{\partial_B \bar{P}}{\bar{P}} + \bar{g}^{rA}(u,x,y) \qquad {\rm if} \; \ba \ne \infty
\ee
\be
 g^{rA} = 2r \bar{Q}^2 h^{AB} \frac{ \partial_B \bar{P}}{\bar{P}} + \bar{g}^{rA}(u,x,y) \qquad {\rm if} \; \ba = \infty
\ee
where $h^{MN}$, $h^{AB}$ are the inverses of $h_{MN}$, $h_{AB}$.

In deriving these results, we used only equations NP1, NP2 and NP3 with vanishing (Ricci and Weyl) curvature components (as is the case for an Einstein spacetime with $\ell^a$ a multiple WAND). The metric deduced above is the most general metric admitting a hypersurface-orthogonal geodesic multiple WAND, for which $\rho_{ij}$ has precisely two distinct eigenvalues, and satisfies $R_{rr}=R_{rM}=R_{rA}=0$.

\subsection{The 5d case}

In 5d, Ref. \cite{Ortaggio:2012hc} proved that $\ba=\infty$ so there are just two cases to consider, $n=2$ and $n=1$. 

For $n=2$ we can perform a coordinate transformation $x \rightarrow x'(x,u)$ to arrange that $h_{MN}(u,x) = \delta_{MN}$ (absorbing a conformal factor into $\bar{P}$). $h_{AB}$ has a single component $h(u,y)$ and $y \rightarrow y'(y,u)$ can be used to set $h=1$. The metric is 
\be
\label{rank2}
ds^2 = -Udu^2 -2 du dr + \frac{r^2}{P^2} \left( d{\bf x} + {\bf A} du \right)^2 + \left[ dy + \left( C + 2r \frac{\partial_y P}{P} \right) du \right]^2
\ee
where we have dropped the bar on $P$. Here bold font denotes objects transforming as vectors in 2d Euclidean space with coordinates ${\bf x}$. The functional dependence is $U=U(u,r,{\bf x},y)$, $P = P(u,{\bf x},y)$, ${\bf A} = {\bf A}(u, {\bf x},y)$, $C=C(u,{\bf x},y)$. 

For $n=1$, $h_{MN}(u,x)$ has a single component which we can set equal to $1$ using $x \rightarrow x'(x,u)$. Although we have $\bar{Q}=1$ in this case, a transformation $y \rightarrow y'(y,u)$ can be used to achieve $h_{AB}(u,y) = Q(u,y)^{-2} \delta_{AB}$ for some function $Q$. The metric is
\be
\label{rank1}
 ds^2 = -Udu^2 - 2 du dr + \frac{r^2}{P^2} \left[ dx + \left( \frac{A}{r^2} + B \right) du \right]^2 + \frac{1}{Q^2} \left[ d{\bf y} + \left( {\bf C} + 2r Q^2 \frac{\nabla P}{ P} \right) du \right]^2
\ee
Again we have dropped the bar on $P$.  Here bold font denotes objects transforming as vectors in 2d Euclidean space with coordinates ${\bf y}$ and $\nabla$ is the gradient operator in this space. The functional dependence is $U=U(u,r,x,{\bf y})$, $P = P(u,x,{\bf y})$, $A = A(u,x,{\bf y})$, $B=B(u,x,{\bf y})$ $C=C(u,x,{\bf y})$, $Q=Q(u,{\bf y})$.  

As mentioned above, these metrics satisfy $R_{rr}=R_{rx}=R_{ry}=0$. 
We will determine the conditions arising from the remaining components of the Einstein equation with cosmological constant:
\be
 R_{ab} = \Lambda g_{ab} 
\ee
The calculations were performed using Mathematica.

\section{5d metric with rank 2 optical matrix}

\label{sec:rank2}

\subsection{The Einstein equation}

For the metric (\ref{rank2}), the components of the Einstein equation in the $x^M,y$ directions give
\be
 R_{My}=0 \qquad \Leftrightarrow \qquad \partial_y A^M = 0
\ee
so ${\bf A} = {\bf A}(u,{\bf x})$
\be
 R_{12} = 0  \qquad \Leftrightarrow \qquad  \partial_{1} A^2 + \partial_{2} A^1 = 0
\ee
\be
 R_{11} - R_{22}=0  \qquad \Leftrightarrow \qquad \partial_{1} A^1 - \partial_{2} A^2 = 0
\ee
These equations are equivalent to (lowering indices with $\delta_{MN}$) 
\be
 \partial_{(M} A_{N)} = \frac{1}{2} \nabla \cdot {\bf A} \delta_{MN}
\ee
i.e. ${\bf A}$ is a conformal Killing vector field of $\delta_{MN}$. This implies that ${\bf A}$ can be eliminated by a change of coordinates $x \rightarrow x'(x,u)$ and absorbing a function of $u,x$ into $P$. So henceforth ${\bf A}=0$. 

We find that
\be
 R_{yy} = \frac{-2\partial_y C}{r} - \frac{4 \partial_y^2 P}{P}
\ee
Hence equating powers of $r$ in $R_{yy}=\Lambda$ gives
\be
 \partial_y C =0 \qquad \Rightarrow \qquad C=C(u,{\bf x})
\ee
\be
\label{Pyy}
 P_{yy} + \frac{\Lambda}{4} P = 0
\ee
where we introduce the notation $P_y = \partial_y P$, $P_{yy} = \partial_y^2 P$ etc. 

The equation $R_{11}+R_{22}  = 2 \Lambda r^2/P^2 $ involves $U$ and $\partial_r U$. It can be integrated with respect to $r$ to give (using (\ref{Pyy}))
\be
 U = -\frac{2m}{r}+\Delta \log P -2r\left[ (\log P)_u - C(\log P)_y \right]+r^{2}\left(3 (\log P)_y^2 -\frac{\Lambda}{4} \right) 
\ee
where $m=m(u,{\bf x},y)$ is the "constant" of integration and $\Delta \equiv P^2 \delta^{MN} \partial_M \partial_N$ is the Laplacian of $h_{MN} = P^{-2} \delta_{MN}$. Note that the $r$-dependence of the metric is now fully determined.

Using these results, we find that the $ur$ component of the Einstein equation is satisfied. Next we consider the $uM$ components. These have terms that are ${\cal O}(1/r^2)$ and ${\cal O}(r)$. The former give
\be
 \partial_M m = 0 \qquad \Rightarrow \qquad m=m(u,y)
\ee
The ${\cal O}(r)$ terms vanish as a consequence of (\ref{Pyy}). The $uy$ component of the Einstein equation has terms that are ${\cal O}(1/r^2)$, ${\cal O}(1)$ and ${\cal O}(r)$. Using (\ref{Pyy}), only the ${\cal O}(1/r^2)$ part is non-trivial, giving
\be
\label{DeltaC}
  \Delta C= - 2 m_y +  8m (\log P)_y 
\ee
If we divide through by $P^2$ then the LHS is independent of $y$. Taking a $y$-derivative and using (\ref{Pyy}) we obtain
\be
\label{dDeltaC}
 12 m (\log P)_y^2 - 6 m_y (\log P)_y + m_{yy} +  \Lambda m  = 0
\ee
The final component of the Einstein equation is the $uu$ component. Using the above solution for $U$ and equations (\ref{Pyy}), (\ref{DeltaC}), (\ref{dDeltaC}), we find that this reduces to 
\be
\label{RTeq5d}
 \Delta\Delta{\log P}+12m\left[ (\log P)_{u}- C(\log P)_{y} \right]  -4(m_{u}-Cm_y)  - P^2 (\nabla C)^2=0 
\ee
where $(\nabla C)^2 = \delta^{MN} \partial_M C \partial_N C$.

In summary, the Einstein equation reduces to equations (\ref{Pyy}), (\ref{DeltaC}) and (\ref{RTeq5d}) with $C=C(u,{\bf x})$ and $m=m(u,y)$. 

Before analyzing these further, we give the expression for $\Phi_{ij}$, the boost-weight zero Weyl components \cite{Durkee:2010xq}. Choosing the spatial basis vectors ${\bf m}_2 = (P/r) \partial/\partial x^1$, ${\bf m}_3 = (P/r) \partial/\partial x^2$ and ${\bf m}_4 =  \partial/\partial y$ gives
\be
\label{Phi}
 \Phi_{ij} = \left( 
 \begin{array}{ccc}
 -m/r^3 & 0 & 0 \\ 0 & -m/r^3 & 0 \\ P \partial_1 C/2r^2 & P \partial_2 C/2r^2 & 0
 \end{array}
  \right)
\ee
From this we can read off $\Phi=-2m/r^3$ and $\Phi^A_{ij} \Phi^A_{ij} = P^2 (\nabla C)^2/(8r^4)$. These are invariants of the solution, i.e., independent of how $\ell$ is completed to a basis. $\Phi_{ij}$ vanishes (which implies that the solution is type III or more special) if, and only if, these two scalar invariants vanish. If $\Phi_{ij} \ne 0$ then there is a curvature singularity at $r=0$. 

\subsection{$m \ne 0$}

In this case, (\ref{dDeltaC}) is a quadratic equation for $(\log P)_y$ whose coefficients depend only on $u,y$. Hence $(\log P)_y$ is independent of ${\bf x}$. It follows that $P$ has the form $P = F(u,y) G(u,{\bf x})$. Equation (\ref{Pyy}) gives
\be
\label{Fyy}
 F_{yy} + \frac{\Lambda}{4} F = 0.
\ee
This is trivial to solve. Using the gauge freedom $y \rightarrow y + f(u)$, $C \rightarrow C-f'(u)$ we can bring any solution of this equation to the form $\alpha(u) \tilde{F}(y)$. The function $\alpha$ can be absorbed into $G$. Hence, without loss of generality (and dropping the tilde)
\be
\label{PFG}
 P = F(y) G(u,{\bf x}).
\ee
where $F(y)$ obeys (\ref{Fyy}). Plugging into equation (\ref{DeltaC}) now gives
\be
 F^2 \left( \frac{m}{F^4} \right)_y = - \frac{1}{2} G^2 \nabla^2 C
\ee
where $\nabla^2 = \delta^{MN} \partial_M \partial_N$ (so $\Delta = P^2 \nabla^2$). The LHS depends only on $u,y$, the RHS depends only on $u,x$ hence we must have
\be
\label{jdef}
 G^2 \nabla^2 C = -j(u)
\ee
\be
F^2 \left( \frac{m}{F^4} \right)_y = \frac{1}{2} j(u)
\ee
for some function $j(u)$. Integrating now gives
\be
 m(u,y) = \frac{1}{2} F(y)^4 \left( j(u) H(y) + k(u) \right)
\ee
where
\be
 H(y) = \int F(y)^{-2} dy
\ee
Now consider equation (\ref{RTeq5d}). Dividing by $F^4$ gives
%\bea
% 0 &=& G^2 \nabla^2 \left( G^2 \nabla^2 \log G \right) + 6 (jH+k) \left( \frac{G_u}{G} - C \frac{F_y}{F} \right) -2 \left( j_u H + k_u \right) \nonumber \\ &+& C \left[ 8(jH+k) \frac{F_y}{F} + 2 \frac{j}{F^2} \right] - \frac{G^2}{F^2} (\nabla C)^2 
%\eea
\be
\label{RTeq5da}
 G^2 \nabla^2 \left( G^2 \nabla^2 \log G \right) + (jH+k) \left(6 \frac{G_u}{G} +2 C \frac{F_y}{F} \right) -2 \left( j_u H + k_u \right)  + 2 \frac{jC}{F^2} - \frac{G^2}{F^2} (\nabla C)^2 =0
\ee
Now differentiate with respect to $y$, use (\ref{Fyy}), and multiply by $F^2$ to obtain
\be
\label{RTeq5db}
 0 = j \left( 6\frac{G_u}{G} -2 C \frac{F_y}{F} \right) -2 C(jH+k) \left( F_y^2+ \frac{\Lambda}{4} F^2 \right) -2 j_u + 2 \frac{G^2 F_y}{F} (\nabla C)^2
\ee
Differentiate again with respect to $y$, using (\ref{Fyy}) to obtain
\be
\label{RTeq5dc}
0=-2\frac{G^2}{F^2} \left( F_y^2 + \frac{\Lambda}{4} F^2 \right) (\nabla C)^2
\ee
There are two cases to consider.

{\bf Case 1.} $\nabla C=0$ (equivalently $\Phi^A_{ij}=0$) hence $C=C(u)$. This is the only case if $\Lambda>0$. Equation (\ref{Fyy}) implies that
\be
\label{lambdadef}
 F_y^2 + \frac{\Lambda}{4} F^2 = \frac{\lambda}{3}
\ee
for some constant $\lambda$.From (\ref{jdef}) we have $j(u)=0$. Equation (\ref{RTeq5db}) reduces to
\be
 0 = -\frac{2}{3} C k F^{-2} \lambda
\ee
hence $C=0$ ($k \ne 0$ since we're assuming $m \ne 0$ here). Equation (\ref{RTeq5da}) reduces to
\be
 G^2 \nabla^2 \left( G^2 \nabla^2 \log G \right) + 6 k (\log G)_u -2 k_u = 0
\ee
This is the 4d Robinson-Trautman equation with "mass function" $k/2$. If we now perform a coordinate transformation $r = F(y)^2 R$ then the metric becomes
\be
\label{RTwarp}
 ds^2 = dy^2 + F(y)^2 \left( - V(u,R,{\bf x})  du^2 - 2 du dR + \frac{R^2}{G(u,{\bf x})^2} d{\bf x}^2 \right)
\ee
where
\be
 V = -\frac{k(u)}{R} + G^2 \nabla^2 \log G - 2R (\log G)_u - \frac{\lambda R^2}{3}
\ee
The 5d metric is a warped product of a 4d Robinson-Trautman solution with cosmological constant $\lambda$ with the $y$-direction. The warp factor $F(y)$ is obtained from (\ref{lambdadef}). If $\Lambda>0$ then we must have $\lambda>0$ and, shifting $y$ by a constant, the solution is
\be
 F(y) = \sqrt{\frac{4\lambda}{3\Lambda}} \sin \left( \sqrt{\Lambda} y/2 \right)
\ee
It is easy to understand why this is a solution: we can write the 5d de Sitter metric as
\be
\label{warp}
 ds^2 = dy^2 + F(y)^2 ds_4^2
\ee
where $ds_4^2$ is the 4d de Sitter metric with cosmological constant $\lambda$. The metric $ds_4^2$ can be replaced with any other 4d Einstein spacetime with the same cosmological constant. Our solution corresponds to choosing this 4d spacetime to be a Robinson-Trautman spacetime with cosmological constant. 

For $\Lambda=0$ we have $\lambda=0$ or $\lambda>0$. In the former case, $F$ is constant so we can take $F=1$ and the solution is simply the product of a 4d Ricci-flat RT solution with a flat 5th direction. In the latter case we have $F(y) = \sqrt{\lambda/3} y$ (shifting $y$ by a constant). In this case, the 5d solution corresponds to writing Minkowski space in the form (\ref{warp}) with $ds_4^2$ the 4d de Sitter metric with cosmological constant $\lambda$, then replacing $ds_4^2$ by a 4d RT spacetime with cosmological constant $\lambda$.

If $\Lambda<0$ then we have have $\lambda>0$, $\lambda=0$ or $\lambda<0$ and the solutions are obtained by writing 5d anti-de Sitter space as a warped product (\ref{warp}) where $ds_4^2$ is 4d de Sitter, Minkowski or anti-de Sitter space, then replacing $ds_4^2$ with a 4d RT solution with the same cosmological constant. 

{\bf Case 2.} In this case,
\be
\label{case2}
F_y^2 + (\Lambda/4) F^2 = 0
\ee
This is possible only if $\Lambda \le 0$. We consider $\Lambda=0$ and $\Lambda<0$ separately.
 
{\bf Case 2.1}.  $\Lambda=0$. We can set $F=1$, (absorbing a constant into $G$) and then $H=y$. There are two subcases to consider depending on whether or not $j=0$.

{\bf Case 2.1.1}. $j=0$. Then $m=k(u)/2$ and the metric is
\be
\label{KK}
ds^2 = -U(u,r,{\bf x}) du^2 - 2 du dr + \frac{r^2}{G(u,{\bf x})^2} d{\bf x}^2 + \left( dy + C(u,{\bf x} )du\right)^2
\ee
where
\be
 U = -\frac{k(u)}{r} + G^2 \nabla^2 \log G - 2r (\log G)_u  
\ee
Note that $\partial/\partial y$ is a Killing vector field in (\ref{KK}) and this metric takes the form appropriate to a Kaluza-Klein reduction to 4d. The KK scalar field is constant and the KK Maxwell field is obtained from the $1$-form potential $C(u,{\bf x}) du$. This is a null field, as it must be for consistency with the scalar field equation of motion. Equation (\ref{jdef}) gives the 4d Maxwell equation:
\be
 \nabla^2 C=0,
\ee
and the 4d Einstein equation is (\ref{RTeq5da})
\be
 G^2 \nabla^2 \left( G^2 \nabla^2 \log G \right) + 6 k (\log G)_u -2 k_u = G^2 (\nabla C)^2
 \ee
This is the 4d RT eq with a null Maxwell field. So this 5d solution is simply the KK lift of a 4d RT solution of Einstein-Maxwell theory, with a null Maxwell field. From (\ref{Phi}) we see that this is an example of a solution with a hypersurface-orthogonal multiple WAND and $\Phi^A_{ij}\ne 0$.

{\bf Case 2.1.2}. $j \ne 0$. Note that this implies $\nabla C \ne 0$ and hence $\Phi^A_{ij} \ne 0$. We can use the gauge symmetry $u \rightarrow u(u')$, $r \rightarrow r/\partial_{u'} u$ to rescale $m$ by an arbitrary function of $u$. This can be used to set $j={\rm constant}$. (We could set $j=1$ but it is convenient to allow more general values e.g. so that the limit $j \rightarrow 0$ can be understood.) Equation (\ref{RTeq5db}) reduces to $G_u=0$ and hence
\be
 G = G({\bf x})
\ee
In the analysis above, we fixed the gauge symmetry  $y \rightarrow y + f(u)$, $C \rightarrow C-f'(u)$ by imposing (\ref{PFG}). However, in the present case of constant $F$, this gauge symmetry remains and we can use it to set $k=0$ (since $j \ne 0$) so $m= y/2$. The metric is
\be
\label{newmetric1}
 ds^2 = -U(r,{\bf x},y) du^2 - 2 du dr + \frac{r^2}{G({\bf x})^2} d{\bf x}^2 + \left( dy + C(u,{\bf x} )du\right)^2
\ee
where
\be
 U = - \frac{jy}{r} + G^2 \nabla^2 \log G 
\ee
The Einstein equation reduces to (\ref{jdef}) (with $j={\rm constant}$) and (\ref{RTeq5da}). The latter simplifies to
\be
G^2 \nabla^2 \left( G^2 \nabla^2 \log G \right) = G^2 (\nabla C)^2 - 2jC
\ee
As with all the $m \ne 0$ solutions, this solution cannot be type III or more special (since $\Phi_{ij} \ne 0$). It cannot be type D either: it can be shown that it is impossible to achieve $\Psi'_{ijk}=0$ by a null rotation about $\ell^a$. Since $G$ is independent of $u$, the above equations might force $C$ to be independent of $u$. In this case, $\partial/\partial u$ is a Killing vector field and the metric also admits a homothety: rescaling $r,y,u$ by the same constant rescales the metric by a constant. 

{\bf Case 2.2}. $\Lambda<0$. Write $\Lambda= -4/L^2$, $L>0$. Equation (\ref{case2}) gives $F(y) = e^{y/L}$ where we have absorbed a constant coefficient into $G$ and used the freedom $y \rightarrow -y$, (with $C \rightarrow -C$, $j \rightarrow -j$) to arrange $F'>0$. We can take $H(y) = -(L/2) e^{-2y/L}$. Equation (\ref{RTeq5db}) is
\be
\label{RTeq5db2}
 G^3 \left( \frac{j}{G^3} \right)_u =\frac{1}{L} \left[  G^2 (\nabla C)^2-jC \right]
\ee
Using this, equation (\ref{RTeq5da}) becomes
\be
\label{RTeq5db3}
G^2 \nabla^2 \left( G^2 \nabla^2 \log G \right) + 6 k (\log G)_u -2 k_u = -\frac{2}{L} k C
\ee
If $j=0$ then (\ref{RTeq5db2}) implies $C=C(u)$, which is case 1. So we must have $j \ne 0$ here. 
Under the gauge freedom $y \rightarrow y+f(u)$, $C \rightarrow C-f'(u)$, $G \rightarrow Ge^{-f/L}$ we have
\be
 j \rightarrow e^{2f/L} j, \qquad k \rightarrow e^{4f/L} k
\ee
We also have the gauge freedom $u \rightarrow u(u')$, $r \rightarrow r/\partial_{u'} u$ which rescales $m$ by an arbitrary function of $u$ and hence rescales $j,k$ by the same arbitrary function. These transformations can be used e.g. to set $j=1$ and $k={\rm constant}$. 

A coordinate transformation $r =F(y)^2 R$ brings the metric to the form
\be
\label{newmetric2}
 ds^2 = (dy+ C(u,{\bf x})du)^2 + e^{2y/L} \left( - V(u,R,{\bf x},y) du^2 - 2du dR + \frac{R^2}{G(u,{\bf x})^2} d{\bf x}^2 \right)
\ee 
where
\be
 V = -\frac{1}{R} \left( k(u) - \frac{j(u)L}{2} e^{-2y/L} \right) + G^2 \nabla^2 \log G - 2R \left[ (\log G)_u + \frac{C}{L} \right] 
\ee
The Einstein equation has been reduced to the three equations (\ref{jdef}), (\ref{RTeq5db2}) and (\ref{RTeq5db3}). 

The above metric appears to be asymptotically locally anti-de Sitter as $y \rightarrow \infty$. The metric on the conformal boundary is similar to a 4d RT solution of Einstein-Maxwell theory (without cosmological constant) where the Maxwell potential is $C du$. However equation (\ref{jdef}) shows that $j$ plays the role of a source for $C$. Maybe one can interpret the boundary metric in terms of a RT solution of Einstein-Maxwell theory coupled to null charged matter.

\subsection{$m=0$}

\label{sec:RTzerom}

Recall that $m=0$ is a necessary, although not sufficient, condition for the spacetime to be of type III or more special.
When $m=0$, the Einstein equation reduces to (\ref{Pyy}) along with
\be
 \Delta C=0
\ee
\be
\label{typeNRT}
  \Delta\Delta \log P = P^2 (\nabla C)^2
\ee
These two equations are identical to the equations governing $d=4$ type III (or more special) RT solutions of Einstein-Maxwell theory \cite{exact} (with the Maxwell field obtained from the potential $Cdu$). The difference here is that $P$ has $y$-dependence, determined by (\ref{Pyy}). The latter equation can, of course, be solved explicitly e.g. $P = \alpha(u,{\bf x}) + \beta(u,{\bf x}) y$ when $\Lambda=0$. Recall that $C$ is independent of $y$.

From (\ref{Phi}), the (5d) solution is type III or more special if, and only if, $\nabla C=0$. In this case, $C=C(u)$. Using the gauge freedom $y \rightarrow y + f(u)$, $C \rightarrow C-f'(u)$ so we can set $C=0$. In this case the above pair of equations reduces to $\Delta \Delta \log P = 0$, which is the same as the equation governing $d=4$ type III (or more special) vacuum RT solutions \cite{exact}. Again, there is a difference in that $P$ has $y$-dependence, restricted by (\ref{Pyy}). 

With $C=0$, a calculation of $\Psi'_{ijk}$ reveals that the solution is type N (or O) if, and only if, we have
\be
\label{typeN}
 \nabla \left( \Delta \log P \right) = 0 \qquad \Rightarrow \qquad \Delta \log P = K(u,y)
\ee
for some function $K$ and
\be
 \partial_y \left( \frac{\partial^2 P}{P} \right)  = 0
\ee
where we have defined a complex coordinate $\zeta =(x^1 + i x^2)/\sqrt{2}$ and $\partial \equiv \partial/\partial \zeta$. Equation (\ref{typeN}) is the same condition as arises for a 4d vacuum RT solution to be type N (or O) \cite{exact}. It implies that 2-surfaces of constant $u,r,y$ have constant curvature.

Let us return to the most general $m=0$ case and attempt to solve the equations by an {\it Ansatz} motivated by our results for $m \ne 0$:
\be
 P(u,{\bf x},y) = F(u,y) G(u,{\bf x})
\ee 
so (\ref{Pyy}) becomes (\ref{Fyy}). Equation (\ref{typeNRT}) implies either $\partial_y F = 0$ or $\nabla C=0$. In the first case we must have $\Lambda=0$ and the solution is the same as case 2.1.1 above but with $m=0$ ($k=0$), i.e., the KK lift of a 4d RT solution of Einstein-Maxwell theory with a null Maxwell field. In the second case the solution is type III (or more special) and, as explained above, we could use the freedom $y \rightarrow y + f(u)$ to set $C=0$. However, we will keep $C(u) \ne 0$ and instead use this freedom as we did below (\ref{Fyy}) to arrange that $F=F(y)$ where $F$ satisfies (\ref{lambdadef}) for some constant $\lambda$, as in case 1 above. A change of coordinates $r=F^2 R$ then brings the metric to the form
\be
 ds^2 = (dy+C(u)du)^2 + F(y)^2 \left( -V(u,R,{\bf x},y) du^2 - 2 du dR + \frac{R^2}{G(u,{\bf x})^2} d{\bf x}^2 \right)
\ee
where
\be
 V = G^2 \nabla^2 \log G - 2 R \left[ ( \log G)_u + C (\log F)_y \right] - \frac{\lambda R^2}{3} 
 \ee
and the single equation of motion is
\be
 \nabla^2 \left( G^2 \nabla^2 \log G \right)=0
\ee
When $C(u)=0$ this solution corresponds to setting $m=0$ ($k=0$) in the warped product solutions of case 1 above.

\section{5d metric with rank 1 optical matrix}

\label{sec:rank1}

\subsection{The Einstein equation}

Now consider the metric (\ref{rank1}). As mentioned above, the $rr$, $rx$ and $ry$ components of the Einstein equation already are satisfied. The $xy^D$ components are
\be
 0 = R_{xD} = -\frac{3r}{2P^2} \partial_D B + \frac{1}{2r} \left( - \frac{1}{P^2} \partial_D A + \frac{4A}{P^3} \partial_D P + \frac{1}{Q^2} \partial_x C_D \right)
\ee
The ${\cal O}(r)$ term gives $\partial_D B = 0$ hence $B=B(u,x)$. But this implies that we can use a change of coordinate $x=x(x',u)$ to set
\be
 B=0.
\ee
The ${\cal O}(1/r)$ terms give ($C_B = \delta_{BA} C^A = C^B$)
\be
\label{CAeq}
 \partial_x C_B = P^2 Q^2 \partial_B \left( \frac{A}{P^4} \right)
\ee
The $AB$ components of the Einstein equation can be treated similarly. They are equivalent to the equations
\be
\label{Ceq}
 \partial_{(A} C_{B)} = -\frac{1}{Q} D_u Q \delta_{AB}
\ee
\be
\label{PQeq}
 \partial_{(A} \left( Q^2 \partial_{B)} P \right) = \left[Q\nabla P \cdot \nabla Q+  \frac{P}{3}\left(Q^2 \nabla^2 \log Q - \Lambda \right)   \right] \delta_{AB}
\ee
where $\nabla$ is the gradient with respect to $y^A$ and $\cdot$ denotes the scalar product defined by $\delta_{AB}$. We have also introduced a modified derivative:
\be
 D_u = \partial_u - {\bf C} \cdot \nabla
\ee
These equations show that ${\bf C}$ and $Q^2 \nabla P$ are conformal Killing vector fields of $\delta_{AB}$. However, we cannot remove ${\bf C}$ by a coordinate transformation of the form $y=y(y',u)$ because ${\bf C}$ depends on $x$. 

The trace of (\ref{PQeq}) gives
\be
\label{tracePQeq}
 \frac{\nabla^2 P}{P}  = \frac{2}{3} \left( \nabla^2 \log Q - \frac{\Lambda}{Q^2} \right)
\ee
Next we consider the $xx$ component of the Einstein equation. This involves $U$ and $U_r$ with the $r$-dependence of all other terms explicit. Integrating gives
\be
 U = \frac{A^2}{P^2 r^2}  + m  - 2r D_u \log P  + r^2 \left( 3 \frac{Q^2}{P^2} (\nabla P)^2 - \frac{Q^2}{2P} \nabla^2 P - \frac{\Lambda}{2} \right) 
\ee
where the arbitrary function $m =m(u,x,{\bf y})$ plays the role of the "constant" of integration.\footnote{t
Note that the $1/r^2$ term in $U$ cancels with another term contributing to $g_{uu}$. This implies that the non-trivial metric components are quadratic or linear functions of $r$, in agreement with Ref. \cite{Pravdova:2008gp}. This reference assumed $\Phi^A_{ij}=0$, which is true for the case considered in this section \cite{Ortaggio:2012hc}.}

Using the above results, we find that the $ur$ component of the Einstein equation is satisfied, and the Ricci scalar obeys $R=5\Lambda$. If we let $E_{ab} = R_{ab} - \Lambda g_{ab}$ then the contracted Bianchi identity now implies
\be
 \partial_r \left( r E_{ux} \right) = \partial_r \left( r E_{uA} \right) = 0
\ee
and hence only the ${\cal O}(1/r)$ terms in the $ux$ and $uA$ components of the Einstein equation can contain new information. The $uA$ component reduces to an equation for $\partial_A m$ which can be integrated to give
\be
 m = M(u,x) P^2 - (\partial_x P)^2 + 2 P \partial_x^2 P
\ee
for some function $M(u,x)$.  Using this in the $ux$ component gives
\be
\label{Pxeq}
 2P \partial_x^3 P + \partial_x \left( MP^2 \right)  = -2P D_u \left( \frac{A}{P^3} \right)
\ee
The remaining component is the $uu$ component. The contracted Bianchi identity reveals that only the ${\cal O}(1/r)$ terms in this can be non-trivial. These terms give:
\be
\label{uueq}
\partial_u M = \ldots
\ee
where the ellipsis denotes a lengthy expression involving $P,Q,A$ and their derivatives.

It is convenient to introduce a complex coordinate $\zeta = (y^1 + i y^2)/\sqrt{2}$ so that $d{\bf y}^2 = 2 d\zeta d\bar{\zeta}$. The $\bar{\zeta}\bar{\zeta}$ component of equation (\ref{Ceq}) gives
\be
 \bar{\partial} C^\zeta = 0
\ee
where $\partial = \partial_\zeta$, and hence
\be
 C^\zeta = W(u,x,\zeta)
\ee
for some holomorphic function $W$. The $\zeta \bar{\zeta}$ component of the same equation gives
\be
\label{duQ}
 \partial_u Q^{-2} = \partial \left( Q^{-2} W \right) + \bar{\partial} \left( Q^{-2} \bar{W} \right)
\ee
Differentiating with respect to $x$ gives
\be
\label{WxQeq}
 \partial \left( Q^{-2} W_x \right) + \bar{\partial} \left( Q^{-2} \bar{W}_x\right)=0
\ee
We also have (\ref{CAeq}) which gives
\be
\label{WxAeq}
 W_x = P^2 Q^2 \bar{\partial} \left( \frac{A}{P^4} \right)
\ee 
together with the complex conjugate of this equation. 

The $\zeta\zeta$ component of equation (\ref{PQeq}) gives
\be
\label{Q2dP}
 \partial \left( Q^2 \partial P \right) = 0
\ee
and hence
\be
\label{Q2dPa}
 Q^2 \partial P = \bar{{\cal F}}
\ee
for some function ${\cal F}(u,x,\zeta)$ depending holomorphically on $\zeta$. 

\subsection{Invariants}

From Ref. \cite{Ortaggio:2012hc}, we know that there exists a basis for which $\Phi_{ij}={\rm diag}(-\Phi,\Phi,\Phi)$. We find that  
\be
\label{Phisol}
 \Phi = 
  \frac{\Lambda}{12}
-  \frac{2}{3}Q^2  \partial \bar{\partial} \log Q 
 \ee
If $\Phi=0$ then $\Phi_{ij}=0$ so the spacetime must be type III, N or O. But a spacetime of type N or (in 5d) type III cannot have a rank one optical matrix \cite{Pravdaetal04}. Hence $\Phi=0$ if, and only if, the solution is conformally flat, i.e., locally isometric to (anti-)de Sitter or Minkowski spacetime. 

Now assume $\Phi \ne 0$ and consider the boost weight $-1$ Weyl components $\Psi'_i$ and $\Psi'_{ijk}$. Use the basis
\be
 {\bf m}_2 = \frac{P}{r} \frac{\partial}{\partial x}, \qquad {\bf m}_3 = Q \frac{\partial}{\partial \zeta}, \qquad {\bf m}_4 = Q \frac{\partial}{\partial \bar{\zeta}}
\ee
It can be shown that the equations derived above (in particular (\ref{WxAeq})) imply that $\Psi'_2 = 0$ and
\be
 \Psi'_3 = \frac{1}{2} rP^4 Q\, \partial \left(P^{-4} \Phi \right)
\ee
with $\Psi'_4$ obtained by complex conjugation. Consider a null rotation about $\ell^a$ with parameters $z_i$. This leaves $\Psi'_2$ invariant (and hence zero) but $\Psi'_3 \rightarrow \Psi'_3 - 2 \Phi z_3$. Hence we can eliminate $\Psi'_3$ by an appropriate choice of $z_3$. Its conjugate $\Psi'_4$ also will vanish. So a null rotation with arbitrary $z_2$ can be used to set $\Psi'_i = 0$. A calculation using the above equations then reveals that, in the new basis, $\Psi'_{ijk}=0$. Hence this null rotation has eliminated all of the boost weight $-1$ Weyl components.

Next we compute $\Omega'_{ij}$ in this new basis. The result is independent of the arbitrary parameter $z_2$ that specifies the basis. Using the above equations (including (\ref{uueq})) we find that the only non-vanishing components are
\be
 \Omega'_{23} = 2 \Phi \left[ r P^{-1} Q \left( P \partial P_x - P_x \partial P \right) + PQ^{-1} \bar{W}_x \right]
\ee
and components related to this by complex conjugation ($\Omega'_{24}$) or symmetry. The solution is type D if, and only if,  $\Phi \ne 0$ and $\Omega'_{ij}=\Psi'_{ijk}=0$. Hence
\be
\label{typeDcond}
  \Phi \ne 0, \qquad P \partial P_x - P_x \partial P =0, \qquad W_x = 0 \qquad \Leftrightarrow \qquad {\rm type \; D}
\ee
If the type D condition is satisfied then we can employ an argument of Ref \cite{Ortaggio:2012hc}. We have noted that the type D condition is satisfied in a basis which, at any point, is specified by an arbitrary parameter $z_2$. We can choose $z_2$ to be an arbitrary function of spacetime position. Since the type D condition is satisfied in this basis, $n^a$ must be a multiple WAND. But the vector field $n^a$ depends  on the arbitrary function $z_2$. $n^a$ will be geodesic only for special choices of function $z_2$, in general it will be non-geodesic. Hence our spacetime admits a non-geodesic multiple WAND. All 5d Einstein spacetimes admitting a non-geodesic multiple WAND were determined in Ref. \cite{Durkee:2009nm}. The possibilities are: $dS_3 \times S^2$, $AdS_3 \times H^2$ and the metric
\be
\label{kkbubble}
            ds^2 = r^2 d\tilde{s}_3^2 + \frac{dr^2}{f(r)} + f(r) dz^2, \qquad  f(r) = k - \frac{m}{r^2} -  \frac{\Lambda}{4}r^2, 
\ee
where $m \ne 0$, $k \in \{1,0,-1\}$, $d\tilde{s}_3^2$ is the metric of a 3d Lorentzian space of constant curvature (i.e. 3d Minkowski or (anti-)de Sitter) with Ricci scalar $6k$ and the coordinate $r$ takes values so that $f>0$. This is an analytically continued version of the Schwarzschild metric (allowing for a cosmological constant).

\subsection{Solving the Einstein equation}

We will show that the only solutions in this class are either conformally flat, or the type D solutions just discussed.

{\bf Case 1.} $\partial \bar{\partial} P = 0$ so
\be
 P = {\cal G}+\bar{\cal G}
\ee
where ${\cal G}(u,x,\zeta)$ is holomorphic. Then we have
\be
 Q^2 \nabla^2 \log Q = \Lambda
\ee
This is Liouville's equation. It says that surfaces of constant $u,r,x$ have constant curvature $\Lambda$. 

\noindent {\bf Case 1.1}. $\Lambda=0$. Using the above equation in (\ref{Phisol}) gives $\Phi=0$ so the solution is flat.

\noindent {\bf Case 1.2}. $\Lambda \ne 0$. The general solution of Liouville's equation is 
\be
 Q^2 = \frac{(1+\Lambda {\cal Z} \bar{\cal Z}/2)^2}{\partial {\cal Z}{\bar{\partial} \bar{\cal Z}}}
\ee
for some function ${\cal Z}(u,\zeta)$. A transformation $\zeta \rightarrow {\cal Z}(\zeta,u)$ can  be used to bring $Q$ to the form
\be
 Q= 1 + \Lambda \zeta \bar{\zeta}/2 
\ee
Using this, equation (\ref{Q2dP}) gives
\be
 \left( 1 + \Lambda \zeta \bar{\zeta} /2 \right) \partial^2 {\cal G} + \Lambda \bar{\zeta} \partial {\cal G} = 0
\ee
which implies $\partial {\cal G} =0$ (e.g. act first with $\bar{\partial}$) hence $\partial P=0$ so $P$ is independent of $\zeta$ (and $\bar{\zeta}$). Now (\ref{WxAeq}) gives  
\be
\label{QWxspecial}
Q^{-2} W_x = \bar{\partial} (A/P^2)
\ee
and plugging this and its complex conjugate into (\ref{WxQeq}) gives $\partial \bar{\partial} (A/P^2) = 0$ hence $A/P^2 = {\cal H} + \bar{\cal H}$ for some ${\cal H}(u,x,\zeta)$. (\ref{QWxspecial}) then gives $W_x = Q^2 \bar{\partial} \bar{\cal H}$. Taking the complex conjugate and acting with $\partial$ gives $\partial (Q^2 \partial {\cal H}) = 0$. This implies $\partial {\cal H}=0$ using the same argument we used to show $\partial {\cal G}=0$. Complex conjugation gives $\bar{\partial} \bar{\cal H}=0$ and hence $W_x=0$. 

These results imply that the solution satisfies the type D condition derived above. Hence it is one of the solutions of Ref. \cite{Durkee:2009nm}. From the fact that it has constant $\Phi$ it follows that it must be $dS_3 \times S^2$ if $\Lambda>0$ or $AdS_3 \times H^2$ if $\Lambda<0$.

\noindent {\bf Case 2.}  $\partial \bar{\partial} P \ne 0$. Let $\chi(u,\zeta,\bar{\zeta})$ be real and satisfy
\be
 \partial \bar{\partial} \chi = Q^{-2}
\ee 
We then have $\partial P = \bar{\cal F} \partial \bar{\partial} \chi = \partial (\bar{\cal F}\bar{\partial} \chi)$ and hence
\be
 P = \bar{\cal F}\bar{\partial} \chi + \bar{\cal H}
\ee  
for some ${\cal H}(u,x,\zeta)$. Equation (\ref{tracePQeq}) is
\be
 \partial \bar{\partial} P = -\frac{1}{3} \left( \partial\bar{\partial} \log Q^{-2} + \Lambda Q^{-2} \right) P
\ee 
Using the above expression for $P$, multiplying by $Q^2/\bar{\cal F}$ and using the definition of $\chi$ gives
\be
\label{tracePQeq2}
 \bar{\partial} \log \partial \bar{\partial} \chi - q \bar{\partial \chi} = q  \frac{\bar{\cal H}}{\bar{\cal F}} -  \frac{\bar{\partial} \bar{\cal F}}{\bar{\cal F}}
\ee
where
\be
 q(u, \zeta ,\bar{\zeta}) = -\frac{1}{3} \left( Q^2 \partial\bar{\partial} \log Q^{-2} + \Lambda \right) 
\ee 
Differentiate with respect to $x$: $\chi$ and $Q$ are independent of $x$ so we obtain
\be
\label{FHeq}
 \left( \frac{\bar{\partial} \bar{\cal F}}{\bar{\cal F}} \right)_x = q \left( \frac{\bar{\cal H}}{\bar{\cal F}} \right)_x
\ee 
Now act with $\partial$ to obtain
\be
 (\partial q) \left( \frac{\bar{\cal H}}{\bar{\cal F}} \right)_x
 = 0
\ee
{\bf Case 2.1}. $\partial q = 0$ hence (as $q$ is real) $\bar{\partial} q = 0$. Act with $\partial$ on (\ref{tracePQeq2})  and recall the definition of $\chi$ to obtain
\be
 \partial \bar{\partial} \log Q^{-2} - q Q^{-2} = 0
\ee
Plugging in the definition of $q$ gives
\be
 8 Q^2 \partial \bar{\partial} \log Q - \Lambda = 0 \qquad \Rightarrow \qquad \Phi = 0
\ee
Hence this case is conformally flat. 

\noindent{\bf Case 2.2}. $({\bar{\cal H}}/{\bar{\cal F}})_x=0$. Hence $\bar{\cal H} = \bar{h} \bar{\cal F}$ for some function $h(u,\zeta)$. From (\ref{FHeq}) we have $(\bar{\partial} \log {\bar F})_x=0$, which implies $\bar{\cal F} = \bar{\cal F}_0 (u,x) \bar{\cal F}_1 (u,\bar{\zeta})$.  
We now have $P = \bar{\cal F} \left( \bar{\partial} \chi + \bar{h} \right)$. But note that the definition of $\chi$ determines $\chi$ only up to addition of a real harmonic function hence we are free to make the redefinition $\chi \rightarrow \chi - g - \bar{g}$ for an arbitrary function $g(u,\zeta)$. We can use this freedom to eliminate $h$. So ${\cal H}=0$ and $P = \bar{\cal F} \bar{\partial} \chi$. Reality of $P$ implies that $\arg {\cal F}_0$ is independent of $x$. By absorbing a phase into ${\cal F}_1$ we can assume that ${\cal F}_0$ is real. 

Now consider the effect of a holomorphic change of coordinates $\zeta = {\cal Z}(u,\zeta')$. In the primed coordinates we have ${\cal F}' = {\cal F}/\partial' {\cal Z}$. Hence by an appropriate choice of ${\cal Z}$ we can arrange that ${\cal F}'_1=1$. Henceforth we drop the primes and subscripts, and so
\be
\label{FPsol}
 {\cal F}=\bar{\cal F}={\cal F}(u,x), \qquad P = {\cal F} \partial \chi
\ee
Reality of $P$ implies that
\be
 \partial \chi = \bar{\partial}\chi
\ee 
Hence $\chi$, $Q$, $P$ are independent of ${\rm Im} \zeta$. Using this, (\ref{WxQeq}) can be rewritten as
\be
\label{lineq1}
 (W_x + \bar{W}_x) \partial \log Q^{-2} + \partial W_x + \bar{\partial} \bar{W}_x = 0
\ee
Act with $\partial \bar{\partial}$ and use holomorphicity of $W$ to obtain
\be
\label{lineq2}
 (W_x + \bar{W}_x) \partial^3 \log Q^{-2} + \left( \partial W_x + \bar{\partial} \bar{W}_x \right)  \partial^2 \log Q^{-2} =0 
\ee
We can view this pair of equations as simultaneous linear equations for $W_x + \bar{W}_x$ and $\partial W_x + \bar{\partial} \bar{W}_x$. 

\noindent {\bf Case 2.2.1}. This linear system has a non-trivial solution. This requires vanishing determinant:
\be
 \left( \partial \log Q^{-2} \right) \left( \partial^2 \log Q^{-2} \right) - \partial^3 \log Q^{-2} = 0
\ee
This can be integrated to give
\be
 \partial^2 \log Q^{-2} = t(u) Q^{-2}
\ee
for some function $t$, which implies
\be
\partial^2 \log \partial^2 \chi = t \partial^2 \chi
\ee 
Integrating gives
\be
 \partial^2 \chi = e^{t \chi +s}
\ee
for some function $s(u)$. Using the above results, equation (\ref{tracePQeq2}) reduces to $t=q$. However, from the definition of $q$ we have $q=-(t+\Lambda)/3$. Hence $q=-\Lambda/4$ which implies $\Phi=0$ so this case is conformally flat.

\noindent {\bf Case 2.2.2}. The linear system (\ref{lineq1}), (\ref{lineq2}) admits only the trivial solution, so $W_x + \bar{W}_x =0$, which implies
\be
 W_x = i \alpha (u,x)
\ee
for some real function $\alpha$. Equation (\ref{WxAeq}) gives
\be
 \bar{\partial} \left( \frac{A}{P^4} \right) = \frac{i\alpha}{P^2 Q^2} 
 \ee
which implies 
\be
 \partial \bar{\partial} \left( \frac{A}{P^4} \right) = i \alpha \partial \left( P^{-2} Q^{-2} \right)
\ee
But note that $\alpha \partial \left( P^{-2} Q^{-2} \right)$ is real so reality of the LHS implies
\be
 \alpha \partial \left( P^{-2} Q^{-2} \right) = 0
\ee

\noindent {\bf Case 2.2.2.1}.  $\partial \left( P^{-2} Q^{-2} \right)=0$. Writing $P,Q$ in terms of $\chi$ gives $\partial^2 (\partial \chi)^{-1} =0$ hence
\be
 \partial \chi = \frac{1}{\gamma (\zeta + \bar{\zeta}) + \delta}
\ee
where $\gamma(u)$ and $\delta(u)$ are real. One can now determine $P,Q$ and plug into (\ref{tracePQeq2}). The final result gives $\Phi=0$, i.e., a conformally flat solution.

\noindent {\bf Case 2.2.2.2}. $\alpha=0$ so $W_x=0$. Together with (\ref{FPsol}), this implies that the type D condition (\ref{typeDcond}) is satisfied if $\Phi \ne 0$. So in this case, the solution is either type D or conformally flat. As explained above, if it is type D then it is one of the solutions found in Ref. \cite{Durkee:2009nm}, i.e., $dS_3 \times S^2$, $adS_3 \times H^2$ or the solution (\ref{kkbubble}). 

\section{A 6d case}

Ref. \cite{Ortaggio:2012cp} obtained necessary conditions on the eigenvalues of the optical matrix of a hypersurface-orthogonal multiple WAND in any number of dimensions. We will investigate the possibility, in 6d, of eigenvalues $\{a,a,b,b\}$ with $a,b \ne 0$, $a \ne b$. 

In the notation of section \ref{sec:coords}, take $d=6$, $n=2$ and finite $\bar{a}$ (i.e. $\rho_{ij}$ has two distinct pairs of coincident non-zero eigenvalues). Using coordinate transformations $x \rightarrow x'(x,u)$, $y \rightarrow y'(y,u)$ we can arrange that
\be
 h_{MN} = P(u,x)^{-2} \delta_{MN} \qquad h_{AB} = Q(u,y)^{-2} \delta_{AB}
\ee
Note that $P$ and $Q$ are not the same as $\bar{P}=\bar{a}$ and $\bar{Q}=1$ (of section \ref{sec:coords}). The metric can be written
\bea
 ds^2 &=& -g^{rr} du^2 - 2 du dr +  \frac{r^2}{\bar{a}(u,x,y)^2 P(u,x)^2} \left( d{\bf x} + {\bf A}(u,x,y) du \right)^2 \\ &+&  \left( 1 +  \frac{r}{\bar{a}} \right)^2 Q(u,y)^{-2} \left[ d{\bf y} + \left( {\bf C}(u,x,y) - Q^{2} (1+r/\bar{a})^{-2} \nabla_{\bf y} \bar{a} \right) du \right]^2 \nonumber
\eea
where we are using 2d vector notation ${\bf x} = x^M$, ${\bf A} = A^M$, $\nabla_{\bf y} \bar{a} = \partial_A \bar{a}$ etc. Using Mathematica we now calculate the Ricci tensor via the Ricci identity. We find that the $rr$, $rx$ and $ry$ components vanish. Vanishing of the $xy$ components gives
\be
 A^M = - P^2 \partial_M \bar{a}+D^M(u,x), \qquad C^A = Q^2 \partial_A \bar{a} + E^A(u,y)
\ee
The traceless part of the $xx$ components gives
\be
 \partial_{(M} D_{N)} \propto \delta_{MN}, \qquad \partial_{(M} \left( P^2\partial_{N)} \bar{a} \right) \propto \delta_{MN}
\ee
where $D_M = D^M$. Hence $D^M$ and $P^2 \partial_M \bar{a}$ are conformal Killing vector fields of $\delta_{MN}$. Similarly the traceless part of the $yy$ components reveals that $E^A$ and $Q^2\partial_A \bar{a}$ are conformal Killing vector fields of $\delta_{AB}$. We can now eliminate $D^M$ by a change of coordinates $x^M = x^M (x',u)$ and we can eliminate $E^A$ by $y^A = y^A(y',u)$. So $D^M = E^A = 0$ henceforth.

Next we consider the trace of the $xx$ and $yy$ components of the Ricci tensor (assuming $\Lambda=0$ now). These are both linear in $g^{rr}$ and $\partial_r g^{rr}$ hence we can solve to determine $g^{rr}$ and $\partial_r g^{rr}$. Consistency of the resulting expressions gives the following equations:
\be
R_x + R_y = 0,
\ee
where $R_x$ is the Ricci scalar of $P^{-2} \delta_{MN}$ and $R_y$ the Ricci scalar of $Q^{-2} \delta_{AB}$. This implies that $R_x$ and $R_y$ are both functions only of $u$ hence we have 2-spaces of constant curvature (of opposite sign).  We also have
\be
 -2\frac{\partial_u Q}{Q} = P^2 \nabla_{\bf x}^2 \bar{a} + \bar{a} R_x
\ee
\be
2 \frac{\partial_u P}{P} = Q^2 \nabla_{\bf y}^2 \bar{a} + \bar{a} R_y
\ee
Calculating the Weyl tensor using Mathematica and making use of the above equations, we find that the boost weight zero Weyl components all vanish so the solution is type III or more special. However, a type III or N solution with hypersurface orthogonal multiple WAND cannot have a rank 4 optical matrix \cite{Pravdaetal04}. It follows that the solution must be conformally flat. 

\section*{Acknowledgments}

This work was supported by a Royal Society University Research Fellowship, European Research Council grant no. ERC-2011-StG 279363-HiDGR, and by a bursary from the Royal Astronomical Society.

\end{document}